\documentclass{aa}

\usepackage{graphicx}
\usepackage{txfonts}
\begin{document}

   \title{Reddening maps of the Magellanic Clouds using spectral energy distribution fitting of red giants}

   \author{H. Netzel
          \inst{1}\fnmsep\thanks{henia@netzel.pl}
          \and
          G. Pietrzy\'nski \inst{1}          
          \and 
          M. G\'orski \inst{1}
          \and
          P. Kervella \inst{2,3}
          \and 
          G. Hajdu \inst{1}
          \and
          R. Kudritzki \inst{4}
          \and
          R. Chini \inst{1,5}
          \and
          W. Kiviaho \inst{2,3}
          \and
          B. Zgirski \inst{7}
          \and
          P. Wielg\'orski\inst{1}
          \and
          D. Graczyk \inst{6}    
          \and
          W. Gieren \inst{7}
          }

   \institute{Nicolaus Copernicus Astronomical Centre, Polish Academy of Sciences, Bartycka 18, PL-00-716 Warszawa, Poland
   \and 
   LIRA, Observatoire de Paris, Universit\'e PSL, Sorbonne Universit\'e, Universit\'e Paris Cit\'e, CY Cergy Paris Universit\'e, CNRS, 5 place Jules Janssen, 92195 Meudon, France.
\and
French-Chilean Laboratory for Astronomy, IRL 3386, CNRS and U. de Chile, Casilla 36-D, Santiago, Chile.
\and
Institute for Astronomy, University of Hawaii, 2680 Woodlawn Drive, Honolulu, HI 96822, USA
\and
Ruhr University Bochum, Faculty of Physics and Astronomy, Astronomical Institute (AIRUB), 44780 Bochum, Germany
\and
Centrum Astronomiczne im. Mikołaja Kopernika, Polish Academy of Sciences, Rabia\'nska 8, 87-100 Toru\'n, Poland
\and
   Universidad de Concepci\'on, Departamento de Astronom\'ia, Casilla 160-C, Concepci\'on, Chile
             }

   \date{Received September 15, 1996; accepted March 16, 1997}

   \abstract 
   {Robust reddening maps of the Large and Small Magellanic Clouds (LMC/SMC) are crucial for a wide range of astrophysical studies, including the calibration of the cosmic distance ladder, investigations of stellar populations in low-metallicity environments, and the characterization of interstellar dust properties.}
   {We aim to construct reddening maps of the Magellanic Clouds using spectral energy distribution (SED) fitting, and to investigate the impact of different stellar atmosphere models on the resulting maps.}
   {We combined optical ($ugriz$) photometry from the SMASH survey with near-infrared ($YJK_{\rm s}$) photometry from the VMC survey for red giant branch (RGB) stars. Observed SEDs were matched to synthetic photometry derived from three atmosphere model grids.}
   {Our maps cover 34.5 deg$^2$ of the LMC and 24.5 deg$^2$ of the SMC at 4 arcmin resolution. We find mean reddening values of $E(B-V)=0.076 \pm 0.022 $ mag for the LMC and $0.058 \pm 0.024$ mag for the SMC. We found that employing different atmospheric models results in differences up to 0.03 mag in the mean reddening. Canonical $R_V$ values for the Magellanic Clouds (3.41 for LMC and 2.74 for SMC, \citealt{gordon2003}) provide results consistent with previous studies.}
   {We confirm higher and more structured reddening in the LMC compared to the SMC, with 30~Doradus standing out as the dominant high-reddening region. Our results show that the absolute reddening scale depends on the choice of stellar atmosphere models, while the relative spatial structure of the reddening maps remains stable.}


   \maketitle

\section{Introduction}

The Large Magellanic Cloud (LMC) and the Small Magellanic Cloud (SMC) are irregular satellite galaxies of the Milky Way (MW). The Magellanic Clouds (MCs), located at distances of approximately 50–63 kpc, are of paramount importance for a wide range of astrophysical studies. They serve as nearby laboratories for investigating galaxy structure and evolution through diverse stellar populations \citep[e.g.][]{jacyszyn_cep,jacyszyn_rrl,ripepi_smc}, as well as for testing stellar physics in different environments \citep[e.g.][]{narloch2022,taormina2024,rathour2025}. In particular, the MCs play a critical role in calibrating the cosmic distance ladder. They contain large populations of standard candles, such as classical and type II Cepheids, RR Lyrae stars, providing an essential rung in the calibration process that ultimately constrains the Hubble constant \citep[e.g.][]{macri2015,riess2019}.

Distances to both Clouds have been determined with remarkable precision using a variety of independent methods. For the LMC, these include red clump (RC) stars \citep{lmc_dist_rc_1,lmc_dist_rc_2,lmc_dist_rc_3}, Cepheids \citep{lmc_dist_cep,storm2011}, RR Lyrae variables \citep{zgirski2023}, and eclipsing binaries \citep{lmc_dist_eb_1,lmc_dist_eb_2,lmc_dist_eb_3}. The most precise recent estimate, based on eclipsing binaries and updated surface brightness-color relations, places the LMC at a distance of $49.59 \pm 0.09_{\rm (stat)} \pm 0.54_{\rm (syst)}$ kpc. Similarly, for the SMC, distance estimates have been obtained using RC stars \citep{smc_dist_rc}, RR Lyrae stars \citep{dist_smc_rrl}, type II Cepheids \citep{dist_smc_t2cep}, eclipsing binaries \citep{dist_smc_eb_1,dist_smc_eb_2,dist_smc_eb_3}, and J-region AGB stars \citep[JAGB,][]{dist_smc_jagb}, with the most recent JAGB-based measurement yielding $62.75 \pm 0.35_{\rm (stat)} \pm 1.45_{\rm (syst)}$ kpc.

Reliable distance determinations, however, require accurate correction for the effects of interstellar dust and gas along the line of sight, which cause extinction and reddening, altering the observed brightnesses and colors of stars. Because dust properties and its spatial distribution are highly non-uniform, reddening can vary significantly across different regions. Constructing accurate reddening maps of the MCs is essential to correct photometric measurements and consequently enabling precise distance determinations and robust studies of stellar populations.

Numerous efforts have been made to construct reddening maps for the MCs, employing different tracers and photometric datasets. Early work by \citet{harris1997}, for example, used the Q-method applied to UBVI photometry of O- and B-type main-sequence stars in the LMC, yielding a mean $E(B-V)$ of 0.13 mag after correcting for Galactic foreground extinction. \citet{zaritsky2002_smc,zaritsky2004_lmc} expanded upon this approach by constructing reddening maps for both MCs using UBVI photometry from the Magellanic Clouds Photometric Survey (MCPS). By fitting stellar atmosphere models to multi-band photometric observations of hot (12\,000\,--\,45\,000\,K) and cool (5\,500\,--\,6\,500\,K) stars, they mapped spatial variations in extinction and its dependence on stellar temperature and population.

RC stars have also been extensively employed as reddening tracers. \citet{subramaniam2005} used RC stars from the OGLE-II photometric maps \citep{2000AcA....50..307U} to study the LMC bar region, finding most regions to have $E(V-I) < 0.1$ mag, with higher values concentrated in the eastern side of the bar. \citet{haschke2011} employed OGLE-III data to create reddening maps for both the LMC and SMC using RC stars and fundamental-mode RR Lyrae stars, confirming the generally low reddening levels across the MCs but highlighting significantly higher extinction in the star-forming region of 30 Doradus. \citet{choi2018} used RC stars from the Survey of the MAgellanic Stellar History (SMASH; \citealt{smash}) to generate reddening maps at 10' and 40" resolution and to investigate the three-dimensional structure of the LMC. \citet{gorski2016} constructed empirical reddening maps from OGLE-III RC stars, obtaining mean $E(B-V)$ values of 0.127 mag for the LMC and 0.084 mag for the SMC. More recently, \citet{skowron2021} used OGLE-IV RC stars to produce reddening maps, deriving mean $E(V-I)$ values of $0.100 \pm 0.043$ mag for the LMC and $0.047 \pm 0.025$ mag for the SMC.

Other studies have utilized other standard candles, such as RR Lyrae stars \citep{pejcha.stanek2009,haschke2011} and Cepheids \citep{inno2016,joshi.panchal2019}. Recently, the SED fitting method, initially employed by \citet{zaritsky2002_smc,zaritsky2004_lmc}, has been utilized in a more advanced form that allows to constrain both stellar and dust parameters by \cite{yanchulova2025} and \cite{lindberg2025} who applied the method to selected regions of the LMC and SMC.

In this study, we investigate reddening maps of the LMC and SMC using SED fitting applied to RGB stars. We combine photometry from the SMASH and VMC surveys to cover both optical and near-infrared regimes. The main goal of this paper, besides constructing the reddening maps, is to investigate in more detail how different stellar atmosphere models affect the resultant maps.

The paper is organized as follows: in Sec.~\ref{sec:methods} we describe the data and methodology. In Sec.~\ref{sec:results} we present our results and discuss them in Sec.~\ref{sec:discussion}. In Sec.~\ref{sec:summary} we summarize our findings.

\begin{figure}
    \centering
    \includegraphics[width=\linewidth]{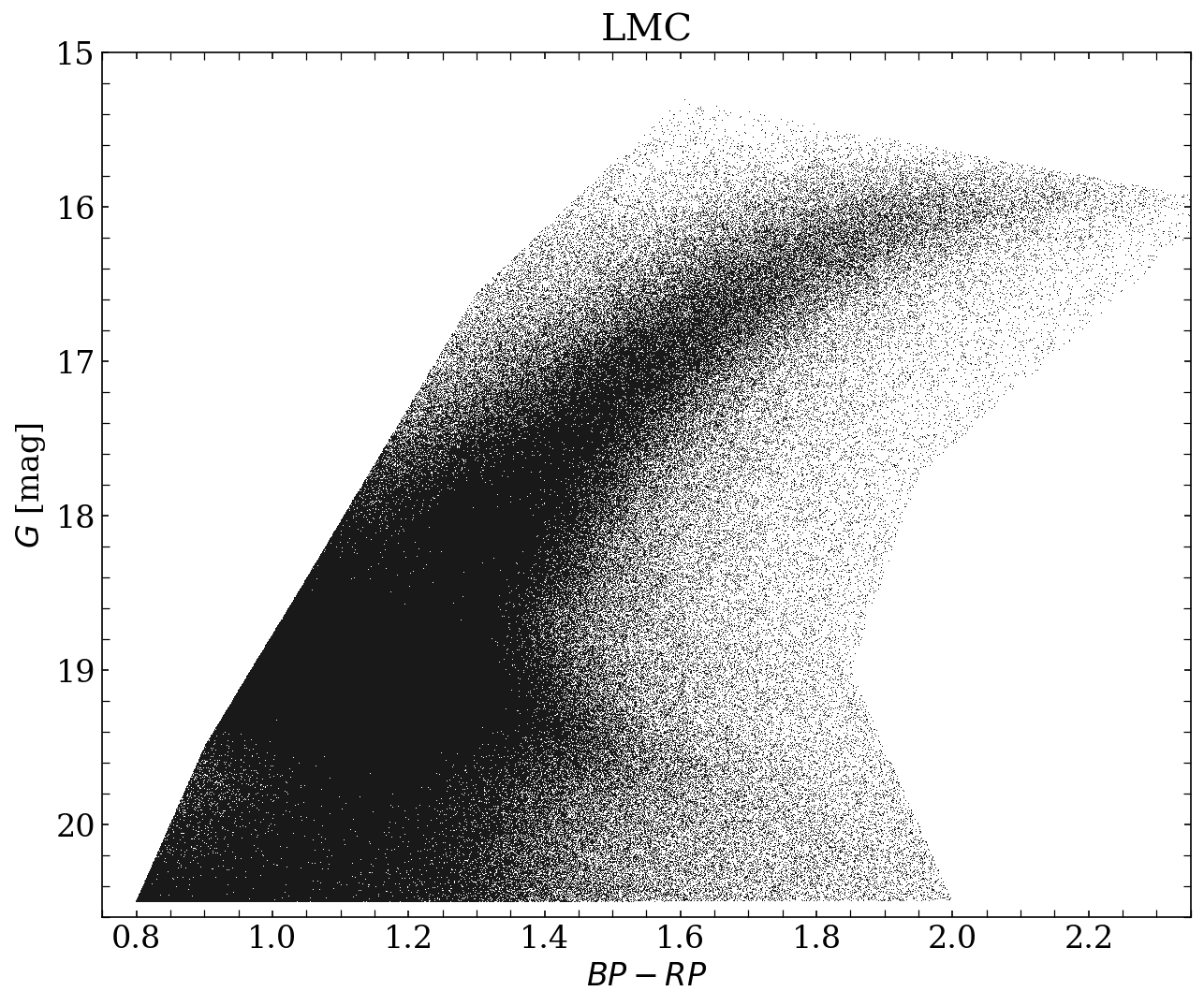}
    \caption{CMD for the LMC with RGB stars selected according to the Gaia EDR3 regions.}
    \label{fig:lmc_cmd}
\end{figure}

\begin{figure}
    \centering
    \includegraphics[width=\linewidth]{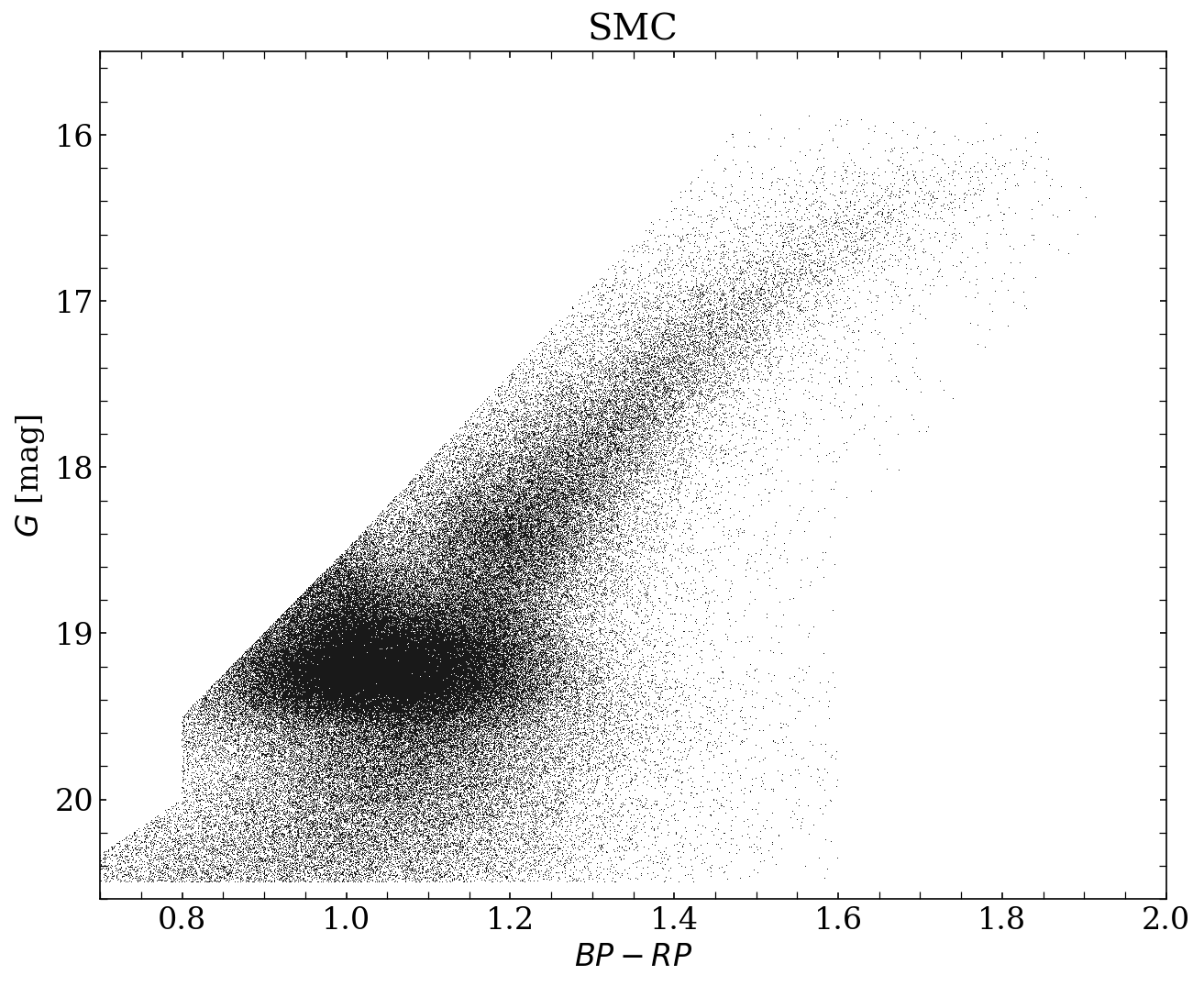}
    \caption{Same as Fig.~\ref{fig:lmc_cmd}, but for the SMC.}
    \label{fig:smc_cmd}
\end{figure}

\section{Methods}\label{sec:methods}

To estimate the reddening affecting individual stars, we performed SED fitting by comparing observed fluxes in eight photometric bands with synthetic photometry derived from model spectra. This approach enables simultaneous estimation of the intrinsic stellar parameters and the reddening due to dust along the line of sight. The photometric data used to construct the observed SEDs are described in Sec.~\ref{subsec:photometry}, while the synthetic photometry is discussed in Sec.~\ref{subsec:synthetic}. 

\begin{figure}
    \centering
    \includegraphics[width=\linewidth]{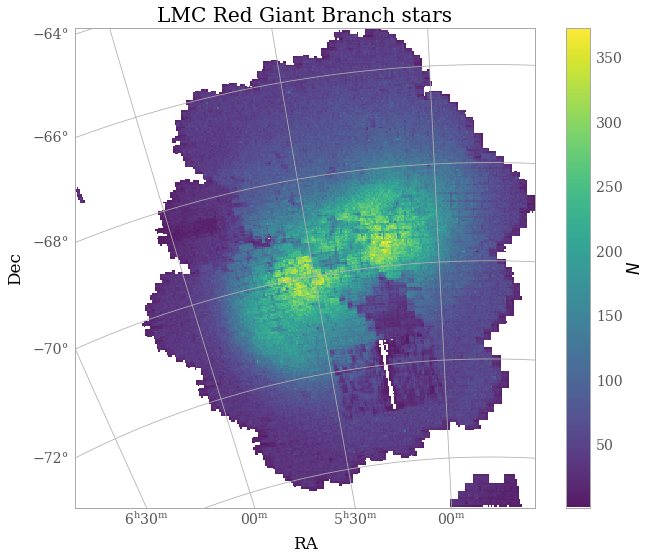}

    \caption{Distribution of RGB stars in the LMC at 2 arcmin resolution.}
    \label{fig:lmc_star_density}
\end{figure}

\begin{figure}
    \centering
    \includegraphics[width=\linewidth]{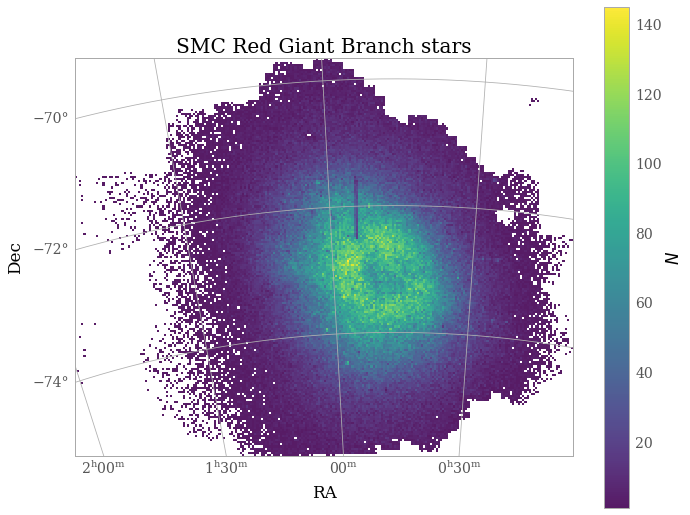}

    \caption{Distribution of RGB stars in the SMC at 2 arcmin resolution.}
    \label{fig:smc_star_density}
\end{figure}

\subsection{Photometric data and sample selection}\label{subsec:photometry}

To build our sample of stars for the LMC and SMC, we combined photometric data from three surveys: the Survey of the Magellanic Stellar History Data Release 2 (SMASH DR2; \citealt{smash}), the Vista Magellanic Cloud Data Release 6 (VMC DR6; \citealt{vmc}), and Gaia Data Release 3 (Gaia DR3; \citealt{Gaia_DR3}).

SMASH is a deep optical survey of the Magellanic System conducted using the DECam instrument \citep{2015AJ....150..150F} on the 4m Blanco telescope. SMASH provides broad spatial coverage of both the central regions and the peripheries of the MCs. Observations were obtained in the SDSS $ugriz$ filters. From this dataset, we selected only those stars that have reliable measurements in all five bands and are located within the LMC or SMC survey fields, i.e. not in the peripheries. Namely, for the LMC we applied a box selection of $RA \in (69.1^\circ, 93.88^\circ)$ and $Dec \in (-75^\circ, -65^\circ)$, resulting in 120\,916\,952 objects. For the SMC, we adopted a box of $RA \in (0^\circ, 30^\circ)$ and $Dec \in (-76^\circ, -69^\circ)$, giving us 40\,278\,610 objects in the SMC fields. Selecting only those stars that have observations in all five bands leaves 53\,165\,335 objects in the LMC and 25\,380\,846 objects in the SMC for further analysis.

To complement the optical coverage, we cross-matched our sample with near-infrared photometry from the VMC survey, which monitored the Magellanic Clouds in the  $YJK_{\rm s}$ bands. All cross-matches in this work were done with the nearest neighbor method adopting a maximum distance of 2 arcsec. Combining optical with near-infrared photometry makes it possible to reduce the degeneracy of SED fitting (see Section~\ref{subsec:fitting}). We used the DR6 catalog \citep{vmc}, adopting aperture-corrected point-source magnitudes. Only stars with photometry in all bands were retained, yielding 23\,577\,023 stars in the LMC and 7\,279\,731 in the SMC in our sample.

We cross-matched our sample with the Gaia DR3 catalog. RGB and RC stars were then identified using the color-magnitude diagram (CMD) selection criteria defined by \citet{luri2021} for Gaia EDR3 data (see their Fig. 2). These CMD-based regions were used as templates to extract candidate RGB stars from our matched multi-band catalog. We note, that because this selection is based solely on broad CMD regions, the resulting sample likely also include stars in nearby evolutionary phases.

Correction for foreground stars is crucial, since nearby stars are less affected by extinction and bias reddening estimates. To address this, we cross-matched our catalog with the foreground classifications of \citet{jimenez-arranz2023_smc, jimenez-arranz2023_lmc}, who used neural networks to separate foreground/background populations of MCs. We adopted a membership probability threshold of 0.98, prioritizing reliability over completeness. After removing the foreground stars, the final sample consists of 4\,135\,193 RGB stars in the LMC, and 583\,083 RGB stars in the SMC.

The final CMDs of the sample of RGB stars in the MCs is presented in Figures~\ref{fig:lmc_cmd} and \ref{fig:smc_cmd}. We note, that the photometric selection applied here is approximate and based solely on wide CMD regions. Consequently, the selected sample includes stars in different evolutionary phases in addition to RGB stars. In particular, the RC sample overlaps with the RGB sample. Moreover, the CMD of the SMC shows an overdensity at around $G\sim 18.5$ mag, and $BP-RP \sim 1.2$ mag that likely corresponds to the AGB clump, while a similar but less pronounced feature can also be seen in the CMD of the LMC.

Figures~\ref{fig:lmc_star_density} and \ref{fig:smc_star_density} show the spatial distribution of our sample in both galaxies. As expected, RGB stars are more numerous towards the central regions of the galaxies, but the most central regions are less populated. This is due to our requirement to select only stars with photometry in all {\it ugriz} bands. Namely, stars in the center generally lack good photometry in the $u$-band.

\subsection{Synthetic photometry}\label{subsec:synthetic}

Synthetic photometry was generated using the Python package \texttt{Astrolib PySynphot}\footnote{\url{https://pysynphot.readthedocs.io/en/latest/index.html}} \citep{pysynphot}, a tool designed for creating synthetic photometry by convolving synthetic spectra with filter transmission curves, with the option of including reddening effects.

We computed fluxes for a grid of stellar parameters representative of RGB stars using three widely used libraries of stellar atmospheres: Kurucz93 \citep[henceforth K93; ][]{k93} Castelli\&Kurucz 2004 \citep[henceforth CK04; ][]{ck04}, and Phoenix \citep{phoenix}. These models were calculated with different treatment of stellar physics and line opacities. Employing different models allows us to explore systematic uncertainties of the SED fitting process. Using \texttt{PySynphot} we interpolated between the individual models to create a uniform grid. We set effective temperatures, $T_{\rm eff}$, to be in the range of 3000 -- 7000 K (step of 100 K) and surface gravity, $\log g$, in the range of 2.0 -- 3.5 dex (step of 0.5 dex). Metallicities, [Fe/H], were chosen based on the average [Fe/H], which was determined to be around $-0.4$ dex for the LMC \citep{choudhury_lmc_feh}, and around $-1.0$ dex for the SMC \citep{choudhury_smc_feh}. We added a margin to these values to reflect any local deviations from the mean. Hence, we set the metallicity range for the LMC from $-0.6$ to 0.0 dex in steps of 0.1 dex, while for the SMC we chose a range of $-1.2$ to $-0.7$ with the same step. The grid in $\log g$ and [Fe/H] is relatively coarse, and these parameters are included primarily to ensure a realistic coverage of stellar atmospheres rather than to derive precise values for individual stars. We note that the selected parameter ranges, in particular for the effective temperature, are slightly broader than typically expected for RGB stars, as we included higher effective temperatures than expected. This choice was intentional, allowing us to test the reliability of the fitting procedure. In particular, to check whether the resulting distribution of effective temperatures is consistent with what is expected for RGB stars.  Not all parameter combinations were available in each model grid. Namely, the $3000-3500$ K range was only provided by Phoenix models. The total number of synthetic flux models generated for the LMC was 343\,728 for a grid based on the K93 and CK04 models, and 391\,468 for a grid based on the Phoenix models. The total number of models having lower metallicity that were calculated for the SMC is 294\,624 for the K93 and CK04, and 335\,544 for the Phoenix models.

Reddening was applied to the synthetic spectra by varying $E(B-V)$ from 0 to 0.3 mag in steps of 0.001 mag, and from 0.3 to 0.5 mag in steps of 0.005 mag. We adopted the \cite{gordon2003} reddening law with $R_V$ values of 3.41 and 2.74, for the LMC and SMC, respectively. Note that part of the reddening toward MCs is due to internal reddening within the MW. We used the maps of \cite{schlegel1998} to estimate the level of MW reddening, which we adopted as the minimum accepted value of $E(B-V)$. We adopted minimum reddening values of $E(B-V)=0.06$ mag and 0.02 mag for the LMC and SMC, respectively. Transmission curves for the $ugrizYJK_{\rm s}$ filters were obtained from the SVO Filter Profile Service \citep{rodrigo2024_svo, svo1, svo2}.

\begin{figure}
    \centering
    \includegraphics[width=\linewidth]{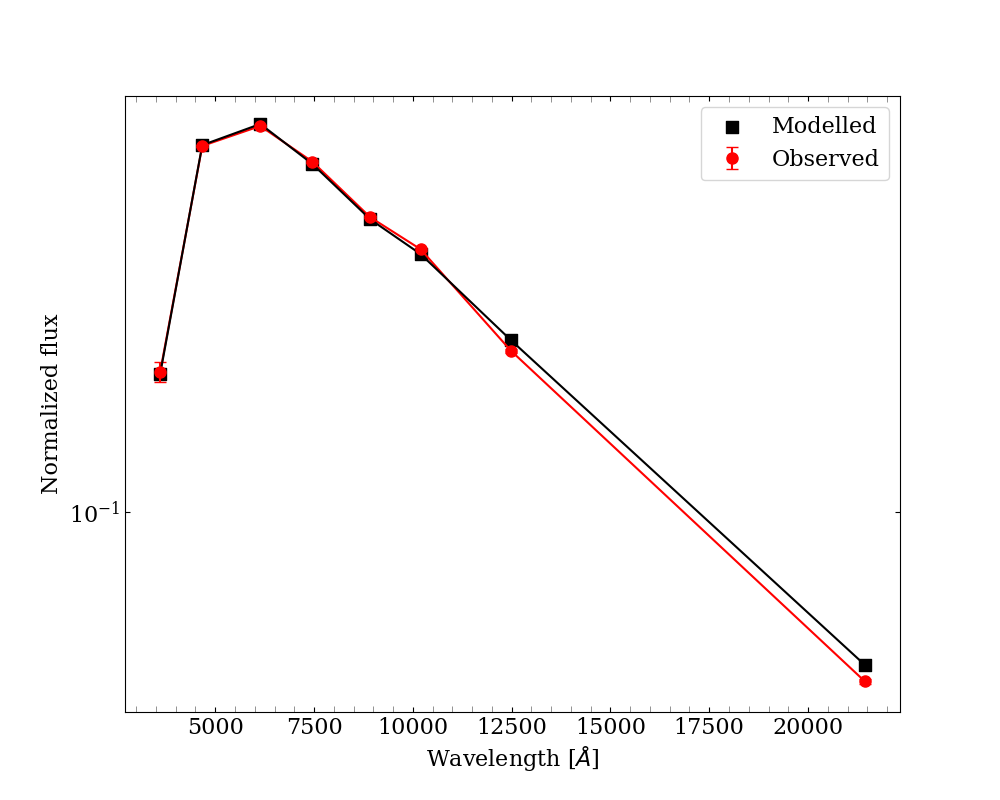}

    \includegraphics[width=\linewidth]{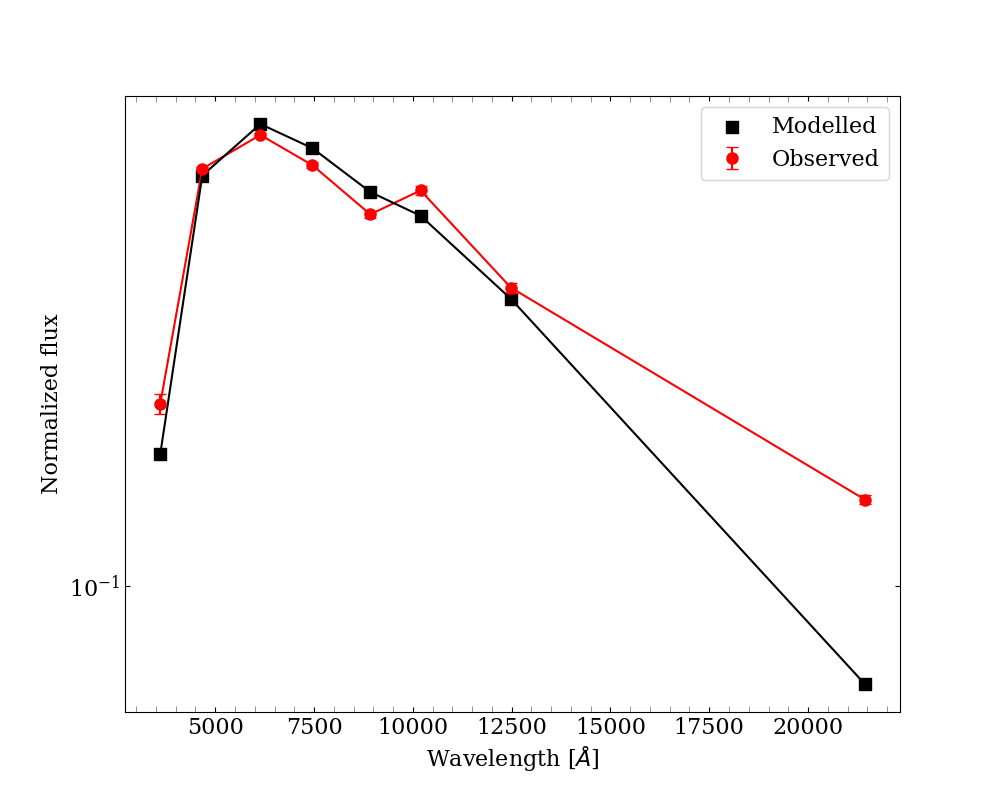}
    \caption{Example of fits for two stars. Observed normalized fluxes are plotted with red line. Calculated fluxes are plotted with black line. Top panel: example of a satisfactory fit for Gaia DR3 4656190370764341504. Bottom panel: example of bad fit for Gaia DR3 4656189408692705792.}
    \label{fig:fit_example}
\end{figure}

\subsection{Fitting procedure}\label{subsec:fitting}

A key limitation of SED fitting is the degeneracy between the effective temperature and extinction. Both of these parameters affect the shape of the SED. While the effective temperature affect the overall slope and the peak of the SED, extinction reddens and dims the SED, particularly in the shorter wavelength range. Using only a limited wavelength range, e.g. only the optical part of the spectrum, restricts the ability to disentangle both effects without any additional information, such as spectroscopic determination of the temperature. In this work we used photometric bands covering optical wavelengths (SMASH $ugriz$) and near-infrared (VMC $YJK_s$), thus allowing us to largely alleviate the degeneracy. While the degeneracy is not completely removed, combining $ugriz$ and $YJK_s$ bands greatly improves the reliability of the derived physical parameters by reducing the overlap in the effects of extinction and temperature on the SED shape. This is due to the fact that different parts of SED are affected differently by extinction and the change in effective temperature. 
Additionally, to address this degeneracy even more, we included a weak prior on $T_{\rm eff}$ described below. 
Observed and synthetic SEDs were normalized by dividing each flux value by the Euclidean norm of the SED vector. This allows for shape comparison independent of the absolute flux values. The fitting of observed and synthetic SEDs was performed using the FAISS library\footnote{\url{https://faiss.ai/index.html}} \citep[Facebook AI Similarity Search;][]{douze2024faiss}, an efficient algorithm designed for fast nearest-neighbor search in high-dimensional spaces. In our context, each observed SED is treated as a vector of flux values, and
the algorithm finds the best-matching synthetic SED by minimizing the squared Euclidean distance between the observed and model SED vectors. This metric corresponds to the sum of squared differences between fluxes in individual filters, and should not be interpreted as a classical $\chi^2$ statistic, since we do not include photometric uncertainties. This approach ensures that the computation time required for the calculation of each map is reasonable and consequently, allowed us to test the differences between maps calculated for various assumptions such as different atmospheric models.

The FAISS algorithm returns the distance between the two vectors. For each star, the algorithm selected the best matching model having the smallest distance within the grid. However, due to the degeneracy between $T_{\rm eff}$ and reddening, choosing only the best-matching model may yield unphysical solutions at the extremes of these parameters. To constrain the solution to physically realistic values, we introduced a weak prior on $T_{\rm eff}$ in the following form:

\begin{equation}
s = \left( \frac{T_{\rm eff} - \mu_{\rm T}}{\sigma_{\rm T}} \right)^2,
\end{equation}
where $s$ quantifies the deviation of the $T_{\rm eff}$ of the model from the expected temperature of RGB stars. Here, $\mu_{\rm T}$ and $\sigma_{\rm T}$ represents the mean and the width of the prior, respectively. We set $\mu_{\rm T} = 4500\, K$  for the LMC and 4700\,K for the SMC, with $\sigma_{\rm T}=500\, K$ for both galaxies. These values are consistent with ranges expected for RGB stars of $\log g$ of 2.0 -- 3.0 dex and reflect the fact that lower-metallicity RGB stars in the SMC, are expected to have typically higher $T_{\rm eff}$ than those in the LMC \citep[e.g.][]{salarisEtal2018}. Among all well fitting models for each star, we chose the one with the most physically plausible temperature, i.e. the model that minimizes $s$. Nevertheless, the best-matching model is not necessarily a well-fitting model, since many objects cannot be adequately represented by our relatively simple grid, which spans a limited number of physical parameters and assumes single, dust-free stars.

In practice, several astrophysical factors can cause observed SEDs to deviate significantly from model predictions. For example, unresolved binarity or multiplicity, the presence of circumstellar material, or variability can all introduce discrepancies. In particular, evolved stars high on the RGB or on the AGB often develop extended dusty envelopes, which result in significant infrared excess. This not only modifies the overall SED shape but also affects the inferred reddening in a way that is not captured by standard extinction laws or used stellar atmosphere models. While the infrared excess could, in principle, be modeled for individual objects using more complex models that account for circumstellar dust, such an approach is beyond the scope of this work. This is due both to the computational cost of fitting large samples and the limited wavelength coverage of our photometric data, which extends only into the near-infrared and does not sufficiently constrain the mid- or far-infrared excess where the infrared excess becomes the most significant.

We visually inspected the fits for a number of stars to arbitrarily determine the limit on the distance provided by FAISS where we consider the best matching models to be satisfactory fits. Example of a satisfactory match (distance of 0.0002) and bad match (distance of 0.008) are presented in Fig.~\ref{fig:fit_example}. We set the limit to be 0.005 and selected only the results fulfilling this criterion for further analysis. 

Visual inspection of the SEDs and photometric data revealed that several fields in the central region of the LMC exhibited instrumental artifacts. The shape of the affected fields matched that of the SMASH CCD chips, further confirming their instrumental origin. After investigating which filters were impacted, we decided to exclude the $r$ and $i$ bands from the fitting process in the central LMC fields.

\begin{figure}
    \centering
    \includegraphics[width=\linewidth]{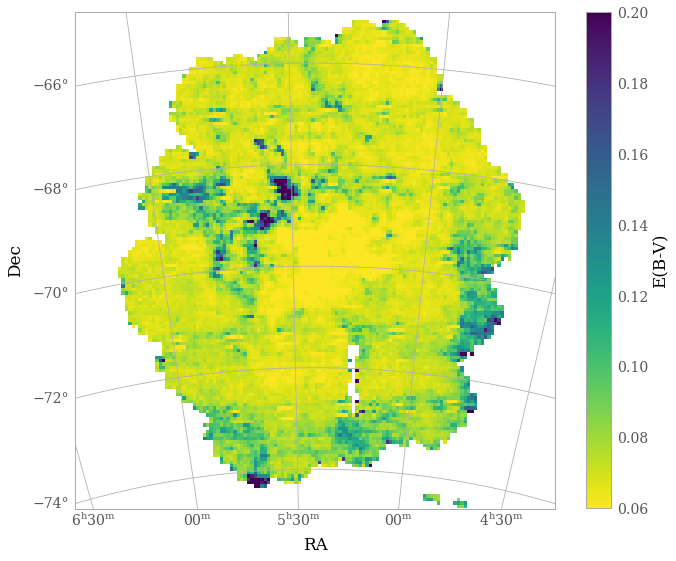}

    \caption{Reddening maps of the LMC calculated for the Phoenix models with a 4 arcmin resolution based on RGB stars.}
    \label{fig:lmc_phoenix_map}
\end{figure}

\begin{figure}
    \centering
    \includegraphics[width=\linewidth]{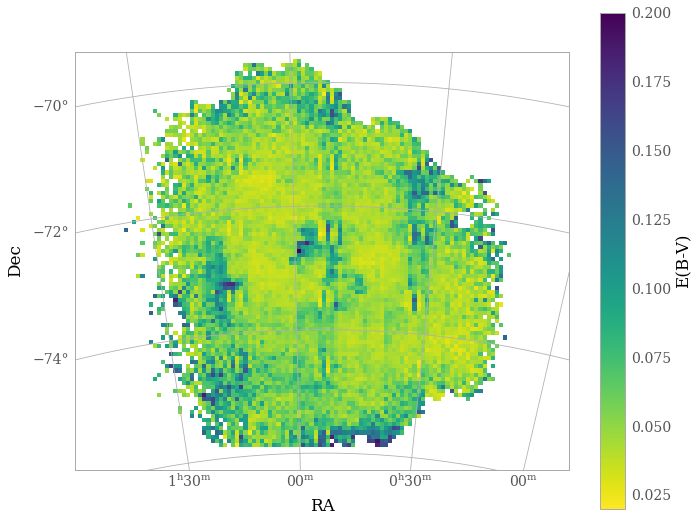}

    \caption{Reddening maps of the SMC calculated for the Phoenix models with a 4 arcmin resolution based on RGB stars.}
    \label{fig:smc_phoenix_map}
\end{figure}

\begin{figure}
    \centering
    \includegraphics[width=\linewidth]{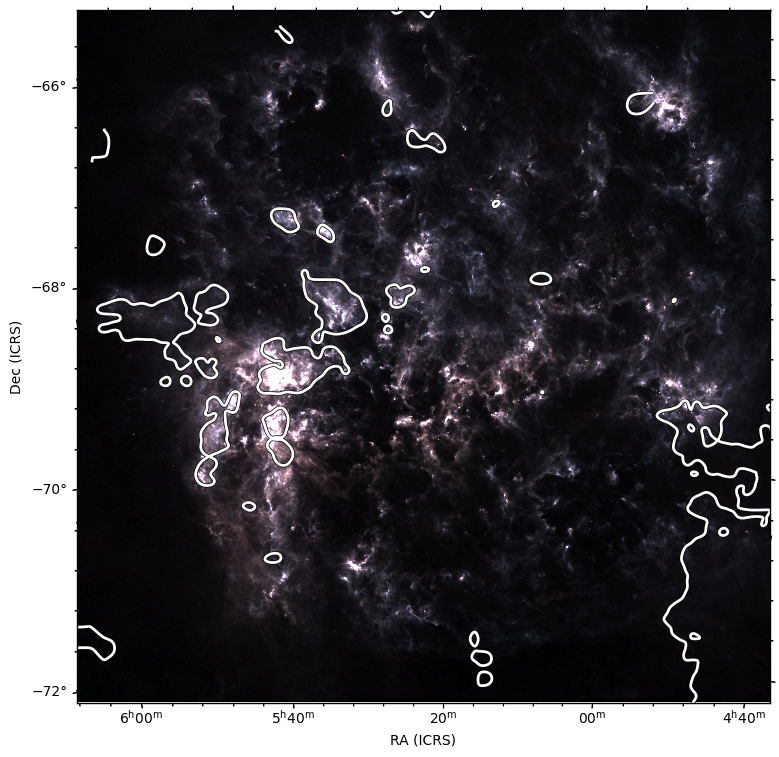}

    \caption{Comparison of dust maps for the LMC from SPIRE/Herschel to $E(B-V)=0.1$ mag contours from the reddening map.}
    \label{fig:lmc_herschel}
\end{figure}

\begin{figure}
    \centering
    \includegraphics[width=\linewidth]{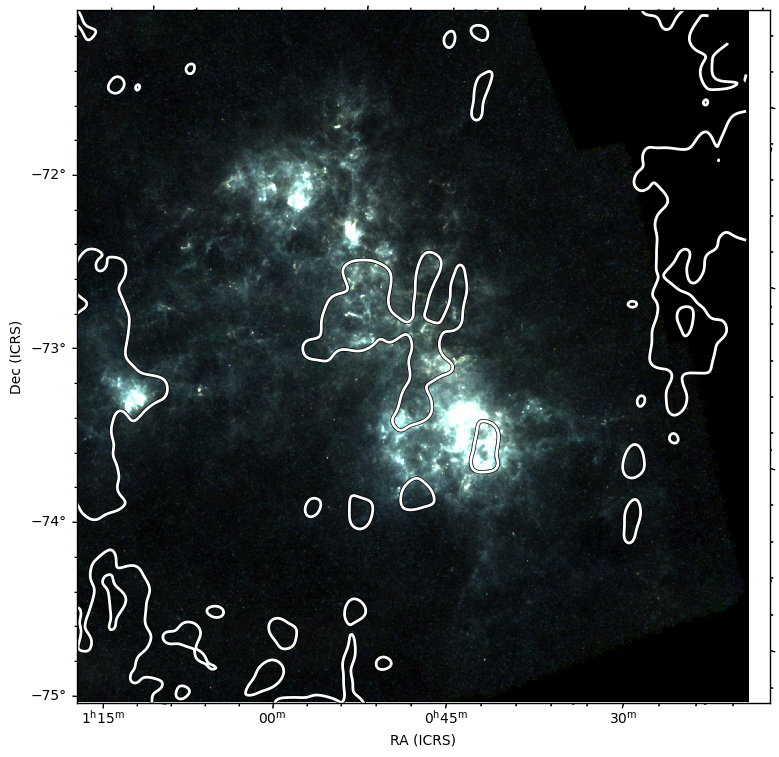}

    \caption{Comparison of dust maps for the SMC from SPIRE/Herschel to $E(B-V)=0.07$ mag contours from the reddening map.}
    \label{fig:smc_herschel}
\end{figure}

\subsection{Reddening maps}

Reddening maps for both MCs, expressed in terms of $E(B-V)$ color excess, were constructed using only stars with satisfactory fits. We calculated several maps based on different atmosphere models. 

Each map was produced at 4 arcmin resolution by binning the $E(B-V)$ values of individual stars and adopting the median per bin to reduce the impact of outliers. Only bins with at least five stars were included. Uncertainties were estimated following the approach of \citet{skowron2021}, using the 16th and 84th percentiles of the distribution.

\section{Results}\label{sec:results}

\subsection{Reddening maps}

We present the reddening maps for the LMC and SMC constructed using the Phoenix models in Figures~\ref{fig:lmc_phoenix_map} and \ref{fig:smc_phoenix_map}, respectively\footnote{These maps are available at \texttt{https://araucaria.camk.edu.pl}.}.

We also calculated reddening maps for the LMC and SMC using the K93 and CK04 models. These maps are included in the Appendix~\ref{sec:other_models} and discussed in more details in Sec.~\ref{sec:dependence_models}. We used the \cite{gordon2003} reddening law with $R_V=3.41$ for the LMC and 2.74 for the SMC.

We note, that in all maps, there are artifacts present that manifest as short double-lined patterns. The spacing and distribution on the pattern corresponds to the size of VMC survey tiles.   
In the map for the LMC in Fig.~\ref{fig:lmc_phoenix_map}, the majority of the LMC and its outskirts are well covered, except for a region at around RA 5h15m00s and Dec --72d15m00s, where RGB stars were not present in our dataset (see Fig.~\ref{fig:lmc_star_density}).

\begin{figure}
    \centering
    \includegraphics[width=\linewidth]{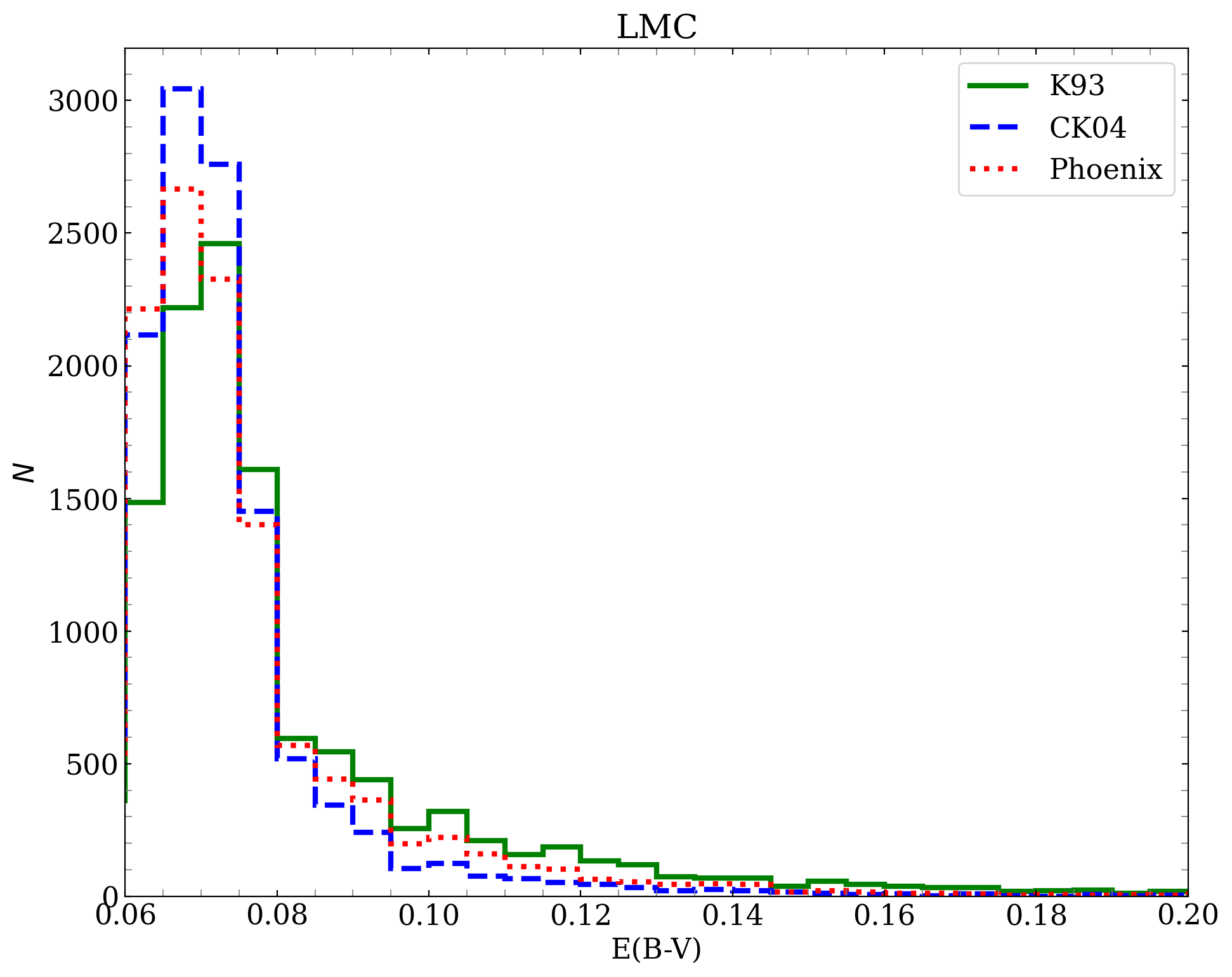}

    \caption{Distribution of reddening in bins for the RGB-based LMC maps calculated using different atmospheric models. Green line: K93 models. Blue dashed line: CK04 models. Red dotted line: Phoenix models.}
    \label{fig:models_lmc_hist}
\end{figure}

The distribution of $E(B-V)$ values for LMC is strongly skewed (Fig.~\ref{fig:models_lmc_hist}), with the majority of bins below 0.1 mag and a long tail extending to 0.405 mag. While the distribution superficially resembles a log-normal form, statistical tests (Kolmogorov-Smirnov, Shapiro-Wilk, and Anderson-Darling) do not support this hypothesis. The mean $E(B-V)$ value for the LMC is $0.076 \pm 0.022$ mag. 

The most characteristic feature of the LMC is the 30 Doradus HII region. It is an active star-forming region that is characterized by larger dust content and consequently it manifests on reddening maps as a high-reddening region. It is clearly visible in the map in this work, consistent with previous studies. Moreover, the 30 Doradus region is clearly visible on all maps, independent of the atmospheric model adopted. Several other darker areas are also visible. We checked whether these areas correspond to dusty star-forming regions. Such regions can be traced as regions of high infrared emission. We used false-color images based on SPIRE/Herschel observations from the HERITAGE project \citep{herschel, herschel_spire, herschel_MC} as a tracer of infrared emission. In Fig.~\ref{fig:lmc_herschel} we show the composite image for the LMC together with contours from our map corresponding to $E(B-V) = 0.1$ mag. The comparison shows that the contours match well not only the 30 Doradus region but also other regions with large dust content, as traced by the prominent infrared emission.

\begin{figure}
    \centering
    \includegraphics[width=\linewidth]{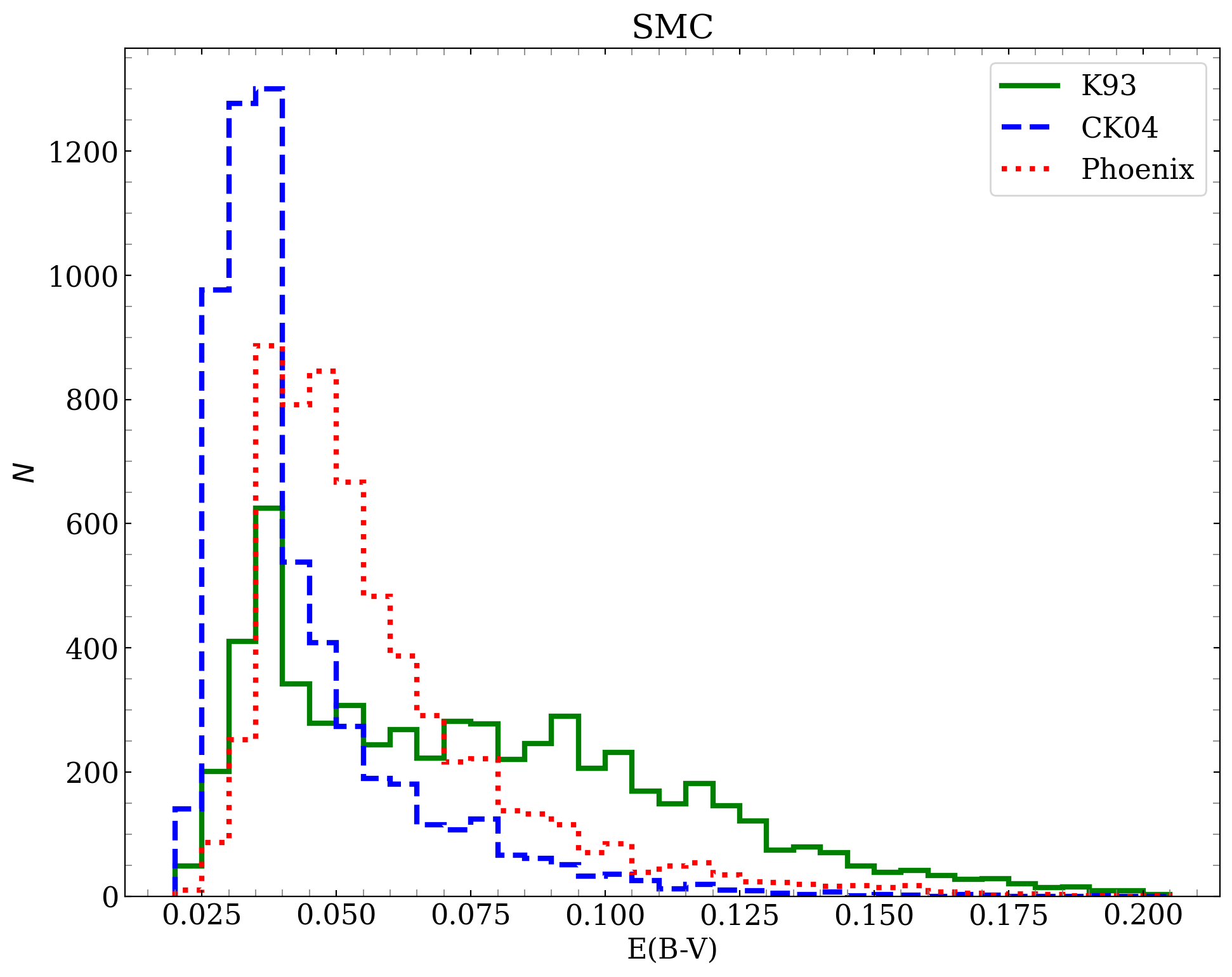}

    \caption{Same as Fig.~\ref{fig:models_lmc_hist}, but for SMC.}
    \label{fig:models_smc_hist}
\end{figure}

The SMC map has on average lower reddening than for the LMC with a mean $E(B-V)=0.058 \pm 0.024$ mag. This is expected, since SMC has lower dust content compared to the LMC \citep{gordon2014}. The distribution of reddening in bins is again skewed (Fig.~\ref{fig:models_smc_hist}) with tail reaching a maximum of 0.204 mag, but it also deviates significantly from the log-normal distribution. In Fig.~\ref{fig:smc_herschel} we plotted the false-color SPIRE/Herschel HERITAGE image together with contours from the SMC map, where $E(B-V) = 0.07$ mag. The contours track some of the most prominent emission regions of the SMC, which includes NGC 456, and NGC 460, as well as the southwest part of the SMC bar. The most prominent emission region in the SMC is N66, connected to the open cluster NGC 346. This area is not highlighted at the $E(B-V)=0.07$ mag level in our map. Moreover, it is also not a high-reddening region in independent maps \citep{gorski2016, skowron2021, chen2022}. The most likely explanation is geometric. Namely, the dust responsible for the infrared emission is located at the far side of the SMC. Interestingly, at the same level of $E(B-V)=0.07$ mag, we identify regions with no detectable infrared emission. This likely indicates dust that is insufficiently heated by young stars, and thus remains faint in the far-infrared, while still producing measurable reddening. Such regions reflect the weaker star-forming activity in the SMC relative to the LMC.

Since RC stars are often used as reddening tracers, we also constructed maps based solely on RC stars from our RGB sample to enable a more direct comparison. The RC stars were selected following the Gaia EDR3 criteria (see Fig. 2 in \citealt{luri2021}). The reddening maps derived from RC stars are consistent with those based on the complete sample. The mean reddening values are $E(B-V)=0.076\pm0.024$ mag for the LMC and $E(B-V)=0.057 \pm 0.026$ mag for the SMC. Consequently, in the following analysis and in comparisons with literature maps, we consider only the RGB-based maps.

\subsection{Dependence on theoretical models}\label{sec:dependence_models}

\begin{figure}
    \centering
    \includegraphics[width=\linewidth]{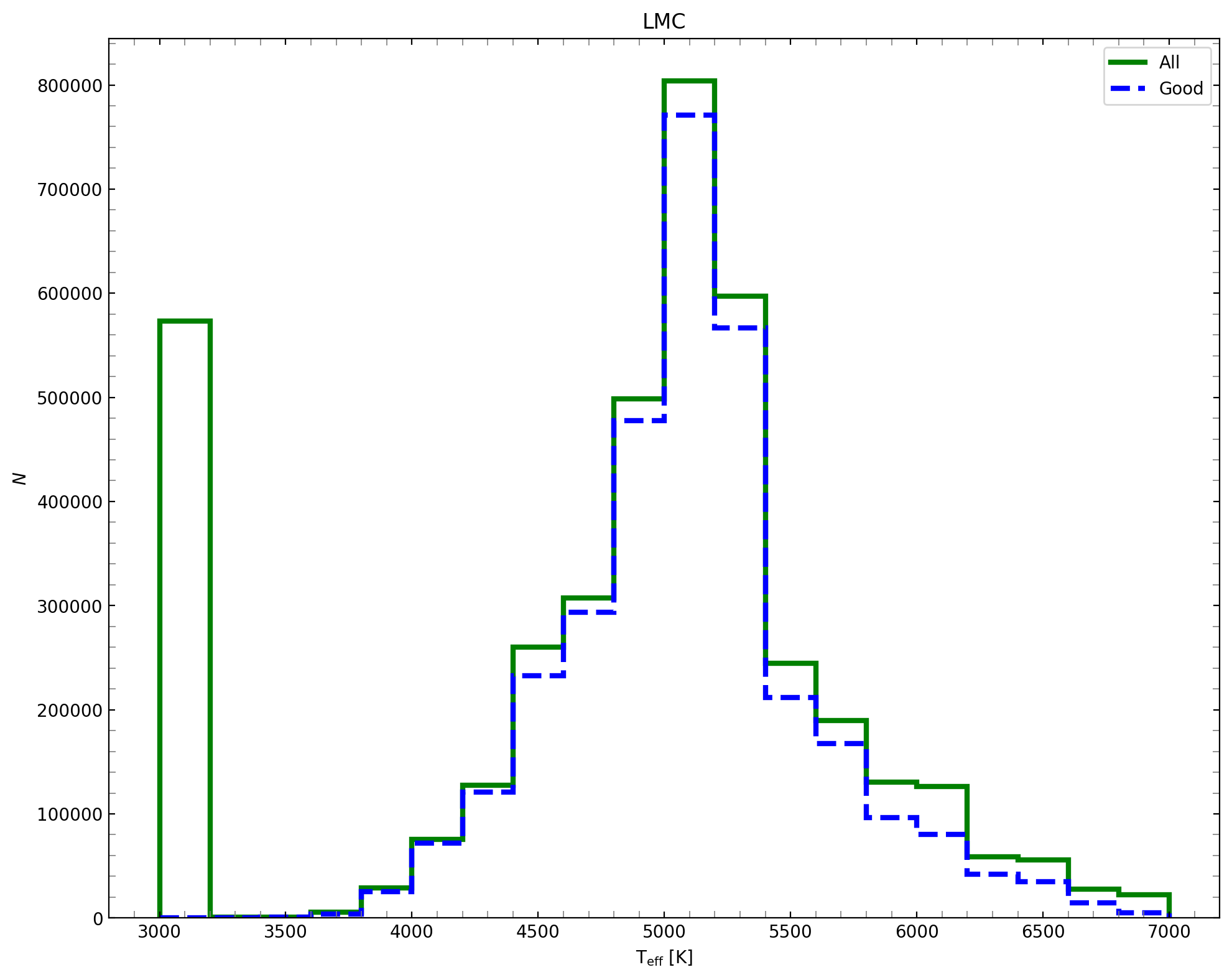}

    \includegraphics[width=\linewidth]{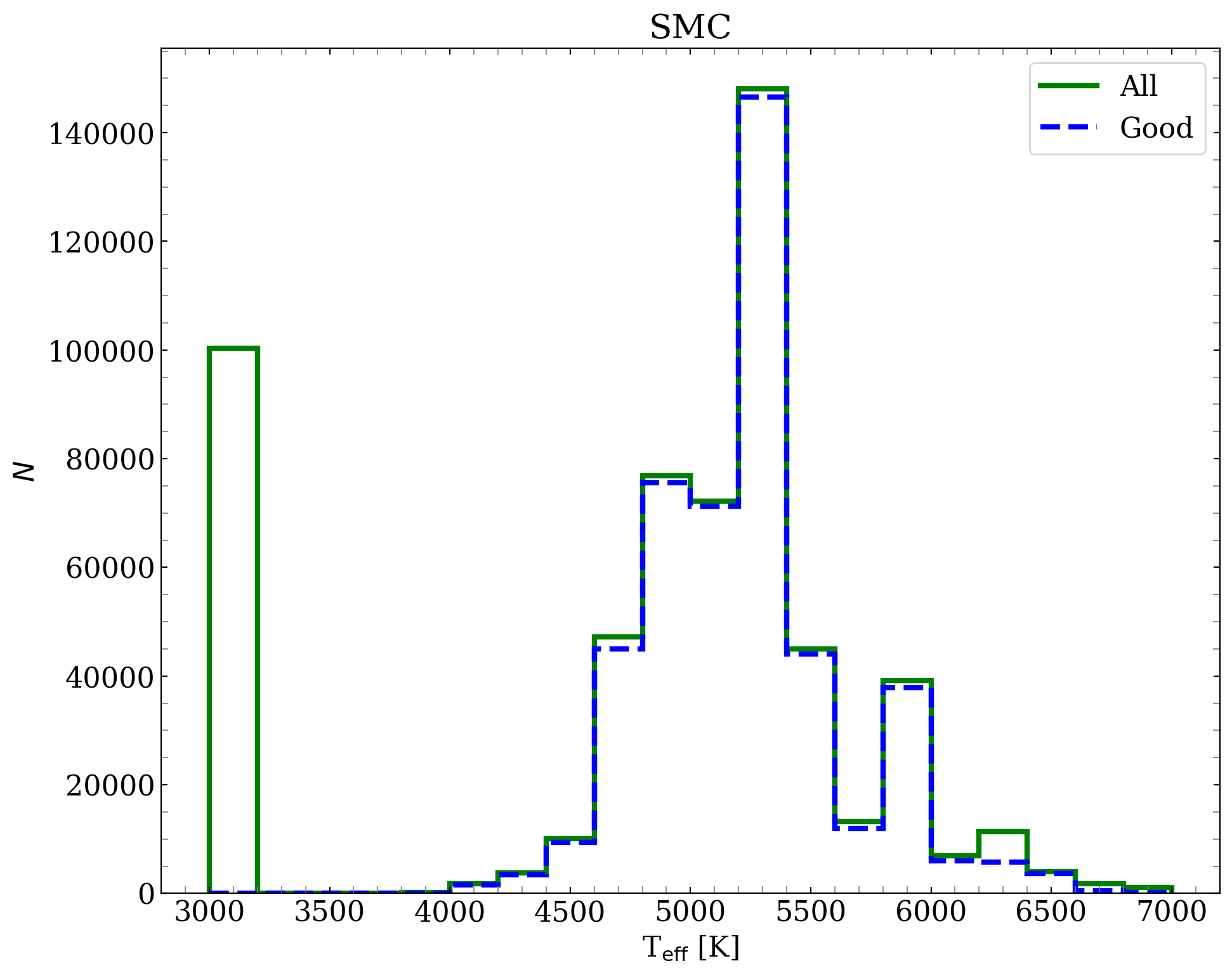}

    \caption{Distribution of effective temperature for modeled RGB stars. Top: LMC. Bottom: SMC. Green solid line corresponds to all stars taken for the modeling. Blue dashed line corresponds to those stars, for which we obtained satisfactory fits (see text for details).}
    \label{fig:teff_rgb_hist}
\end{figure}

We calculated maps using three different grids of atmospheric models: K93, CK04, and Phoenix. The K93 grid calculated with the ATLAS9 code contain over 7000 stellar atmosphere models covering a wide range of effective temperature, metallicity and gravity. The grid was calculated with the opacities of \cite{ag98_opac}. CK04 improved on K93 with updated abundances \citep{gs1998_opac}, revised TiO and H$_2$O line lists, and extended molecular opacities, resulting in significantly improved SEDs for cool stars ($T_{\rm eff} \lesssim 4500$ K). Phoenix models use the abundances of \cite{asplund2009}. Moreover, they use spherically symmetric models in contrast to the plane parallel models of K93 and CK04. The spherical geometry together with updated abundances, line lists and molecular lists for opacities, result in more reliable models for cool RGB stars. Moreover, Phoenix models reach lower effective temperatures than K93 and CK04. We used models down to $T_{\rm eff} = 3000$ K, whereas for K93 and CK04, the grid was limited at $T_{\rm eff} = 3500$ K. In the context of SED modeling of RGB stars, we expect Phoenix models to yield the most realistic results, with CK04 a good second option. Due to the limitations of K93 models in the context of cool stars, we expect them to yield the least reliable results for our sample.

In Appendix~\ref{sec:other_models} we provided LMC and SMC maps calculated using the K93 and CK04 models in Figures~\ref{fig:lmc_k93_ck04}, and \ref{fig:smc_k93_ck04}. Visual comparison between the three maps for the LMC and SMC instantly suggests that the K93-based maps have, on average, larger reddening values than the CK04-based and Phoenix-based maps. However, regardless of the selected models, the characteristic features of the 30 Doradus region in the LMC and the higher-reddening regions of the SMC are well preserved in all maps.

To quantitatively compare reddening calculated with three sets of models, we plotted the $E(B-V)$ distributions for bins from the reddening maps in Figures~\ref{fig:models_lmc_hist} and \ref{fig:models_smc_hist} for LMC and SMC, respectively. For the LMC, the Phoenix-based and CK04-based distributions agree well, while the K93-based distribution is slightly shifted relative to the other two. The mean values are $E(B-V)_{\rm K93} = 0.082$, $E(B-V)_{\rm CK04} = 0.073$, and $E(B-V)_{\rm Phoenix} = 0.076$. A similar trend is also observed for the SMC. Namely, the CK04-based and Phoenix-based distributions are comparable with CK04 showing more low-reddening values, while K93 is again shifted towards higher-reddening values. The mean values are $E(B-V)_{\rm K93} = 0.076$, $E(B-V)_{\rm CK04} = 0.043$, and $E(B-V)_{\rm Phoenix} = 0.058$.

\subsection{Effective temperature}

In addition to $E(B-V)$, the grid of synthetic photometry spanned the parameter space of effective temperature, surface gravity, and metallicity. Because our SED fitting was performed on a population of RGB stars with a known distribution of effective temperatures, the fitted distribution provides a good test of the reliability of the method and of whether the degeneracy between effective temperature and reddening compromised the results. In Fig.~\ref{fig:teff_rgb_hist} we show the distribution of effective temperatures obtained for all RGB  stars in the modeling, and only to those with satisfactory fits.

Selecting only well fitted stars eliminated the high peak at 3000\,K that resulted from stars that likely cannot be modeled within our grid. The distribution of RGB stars peaks at around 5000\,K, with a tail towards lower and higher values from 3500\,K to 7000\,K. The peak of the $T_{\rm eff}$ distribution for SMC is shifted slightly towards higher temperatures than for the LMC, with mean temperature of $T_{\rm eff}=5180$\,K. This is expected given the lower metallicity of the SMC, as RC stars tend to have higher $T_{\rm eff}$ for more metal-poor populations \citep{girardi.salaris2001}. Note that the mean values of distributions for both galaxies are higher than the mean temperature assumed in the prior. This demonstrates that prior did not significantly influenced the fit, but rather served its main purpose of eliminating extreme unphysical values while preserving the underlying temperature distribution of both samples.

\section{Discussion}\label{sec:discussion}

In this study we constructed a collection of reddening maps for the LMC and SMC using SED fitting method. We adopted different atmosphere models to explore their influence on the resultant maps. 
Here we compare our results to maps from the literature.

To compare our results with other reddening maps we calculated differences of the $E(B-V)$ color excess between our map and a given literature map, i.e.:

\[\Delta E(B-V) = E(B-V)_{this\,work} - E(B-V)_{literature}.\]

We analyze these differences both spatially (as maps) and statistically (as histograms). In Sec.~\ref{sec:sed_maps} we compared our results and discuss them in the context of other maps utilizing the SED-fitting methods. In Sec.~\ref{sec:rc_maps} we compared our maps with maps based on the RC colors, since our sample of RGB stars also include RC stars.

\subsection{Comparison with SED-based studies}\label{sec:sed_maps}

\begin{figure}
    \centering
    \includegraphics[width=0.9\linewidth]{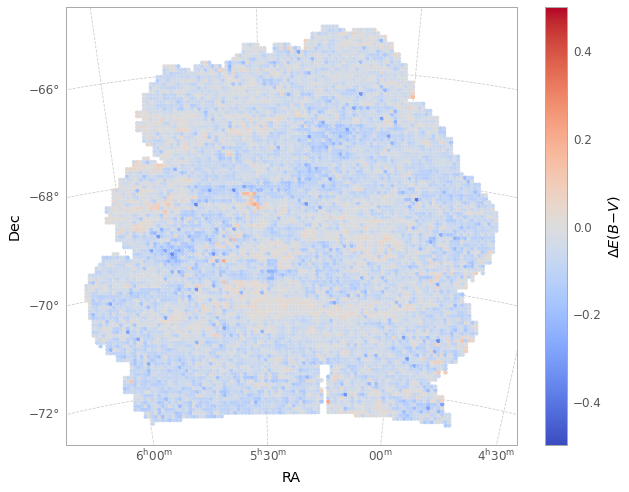}

    \includegraphics[width=0.9\linewidth]{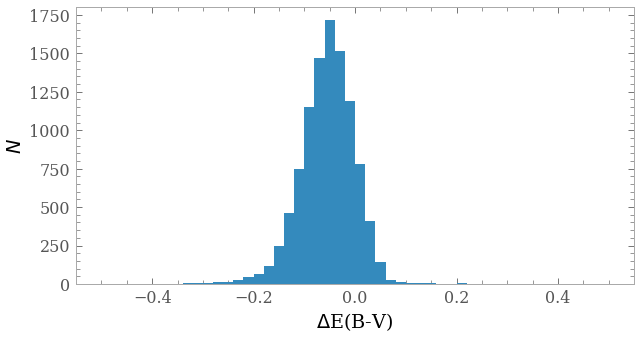}
    \caption{Comparison between our map for the LMC and the map by \cite{zaritsky2004_lmc}. Top: spatial distribution of differences. Bottom: histogram of differences.}
    \label{fig:comp_lmc_zaritsky}
\end{figure}

\begin{figure}
    \centering
    \includegraphics[width=0.9\linewidth]{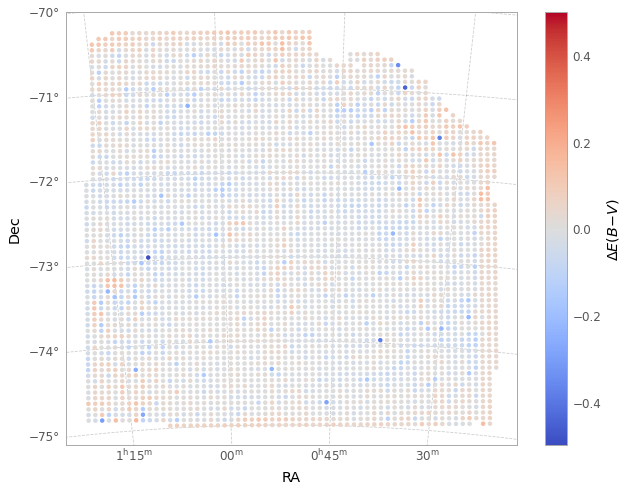}

    \includegraphics[width=0.9\linewidth]{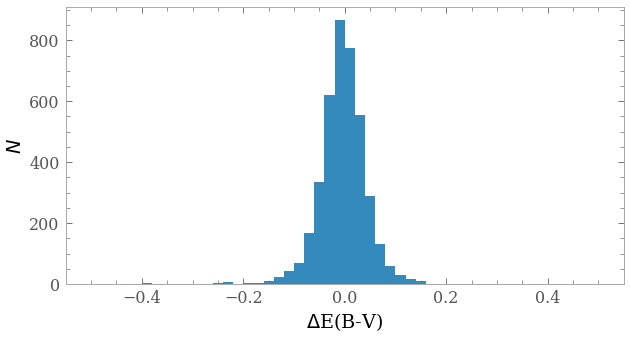}
    \caption{Comparison between our map for the SMC and the map by \cite{zaritsky2002_smc}. Top: spatial distribution of differences. Bottom: histogram of differences.}
    \label{fig:comp_smc_zaritsky}
\end{figure}

\cite{zaritsky2002_smc} and \cite{zaritsky2004_lmc} used the SED fitting method to construct extinction maps of SMC and LMC, respectively, using both hot and cool stars. We compared our maps with their maps based on cool stars. We converted their $A_V$ maps to $E(B-V)$ using $R_V=3.41$. Figures~\ref{fig:comp_lmc_zaritsky} and \ref{fig:comp_smc_zaritsky} show the comparison between their results and our maps. Overall, the mean $\Delta E(B-V)$ is $-0.05 \pm 0.05$ mag. For the SMC, the differences are smaller and more uniform. The mean difference is $-0.004 \pm 0.047$ mag.

\begin{figure}
    \centering
    \includegraphics[width=0.9\linewidth]{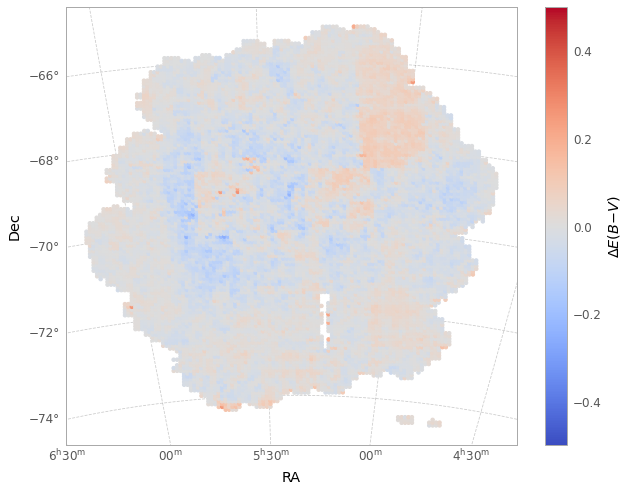}

    \includegraphics[width=0.9\linewidth]{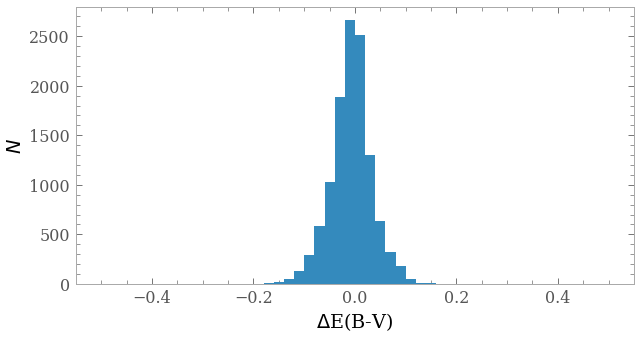}
    \caption{Comparison between our map for the LMC and the map by \cite{chen2022}. Top: spatial distribution of differences. Bottom: histogram of differences.}
    \label{fig:comp_chen_lmc}
\end{figure}

\begin{figure}
    \centering
    \includegraphics[width=0.9\linewidth]{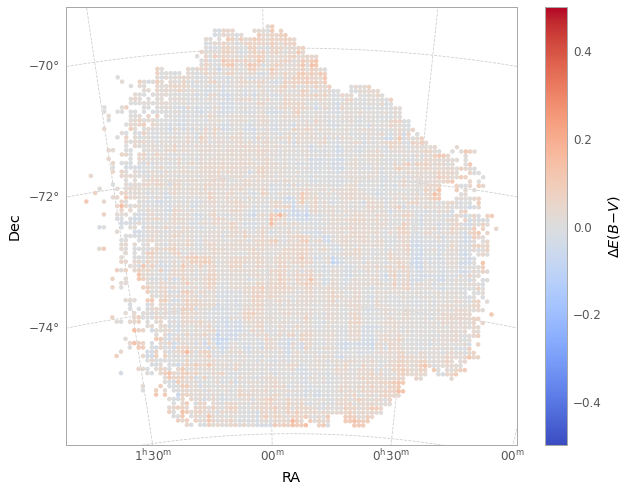}

    \includegraphics[width=0.9\linewidth]{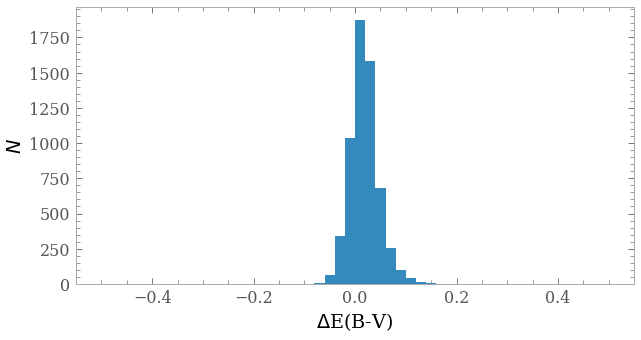}
    \caption{Comparison between our map for the SMC and the map by \cite{chen2022}. Top: spatial distribution of differences. Bottom: histogram of differences.}
    \label{fig:comp_chen_smc}
\end{figure}

\cite{chen2022} applied a SED fitting approach for both MCs using different stellar populations across the CMD. Figures~\ref{fig:comp_chen_lmc} and \ref{fig:comp_chen_smc} show the comparison between their results and our maps. The mean difference between our map and that of \cite{chen2022} is $-0.007 \pm 0.041$ mag for the LMC, and $0.019\pm 0.02$ mag for the SMC. The spatial distribution of these differences shows localized regions with systematically higher or lower values, including sharp, linear boundaries in some areas. These patterns are consistent with unphysical, likely instrumental features present in the \cite{chen2022} LMC and SMC maps.

Recent studies by \cite{lindberg2025} and \cite{yanchulova2025} used SED fitting to derive reddening maps for selected regions in the MCs.

\cite{yanchulova2025} analyzed one $12' \times 6.5'$ field in the south-western bar of the SMC and produced a high-resolution (7") map of $A(V)$. Because of the large difference in spatial resolution and the use of different extinction laws, only the overall level of extinction can be compared. Limiting our map to the same area and converting our $E(B-V)$ values to $A(V)$, we obtain a mean value of $A(V)=0.21$ mag. In comparison, \cite{yanchulova2025} found an $A(V)$ distribution with a peak at about 0.6 mag. This difference is likely related to the local character of the field and to differences in the adopted extinction law.

\cite{lindberg2025} studied 33 fields in the SMC and 56 fields in the LMC, with each field covering 5--7 arcmin$^2$. Because of the sparse spatial coverage and differences in the applied methods, a quantitative comparison with our reddening maps is not straightforward. Their results nevertheless indicate that extinction properties vary across both MCs.

\cite{urbaneja2017} analyzed 90 blue supergiants in the LMC using spectroscopic data and spectral modeling. The reddening values $E(B-V)$ for these stars peak at around $E(B-V)=0.08$ mag, which is consistent with the maximum of the reddening distribution derived in this work. The inferred extinction curves show a large star-to-star variation. However, blue supergiants represent a very young stellar population and are therefore not directly comparable to the RGB stars analyzed in this study. Their reddening may be more strongly affected by local environments.

\subsection{Comparison with RC-based maps}\label{sec:rc_maps}

A number of studies used RC stars as reddening tracers. These stars, as they are confined to a specific region of the CMD, offer a unique way to measure distances and reddening thanks to their narrow intrinsic color range. Here, we compared our maps with the maps calculated by \cite{gorski2016} and \cite{skowron2021} who used OGLE data for RC stars and obtained maps of both MCs. We also compared our results with values from \cite{choi2018}, who used SMASH DR1 data, and utilized RC stars to create a reddening map for LMC.

\begin{figure}
    \centering
    \includegraphics[width=0.9\linewidth]{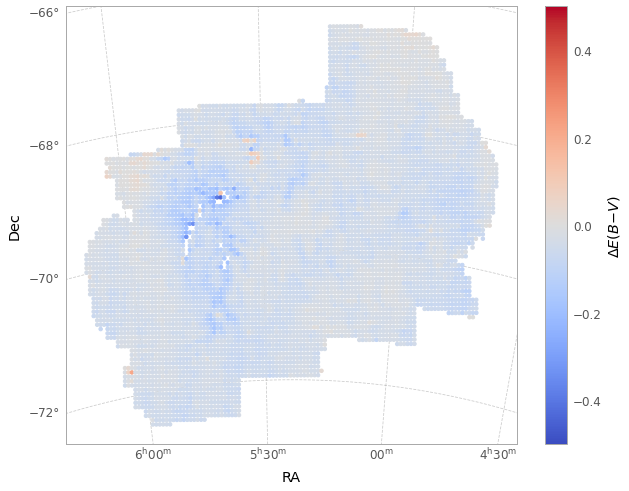}

    \includegraphics[width=0.9\linewidth]{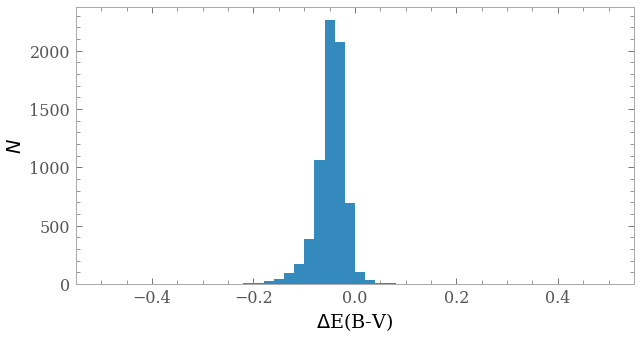}
    \caption{Comparison between our map for the LMC and the map by \cite{gorski2020}, but calibrated only with Na I D1-based reddening. Top: spatial distribution of differences. Bottom: histogram of differences.}
    \label{fig:comp_gorski_lmc}
\end{figure}

\begin{figure}
    \centering
    \includegraphics[width=0.9\linewidth]{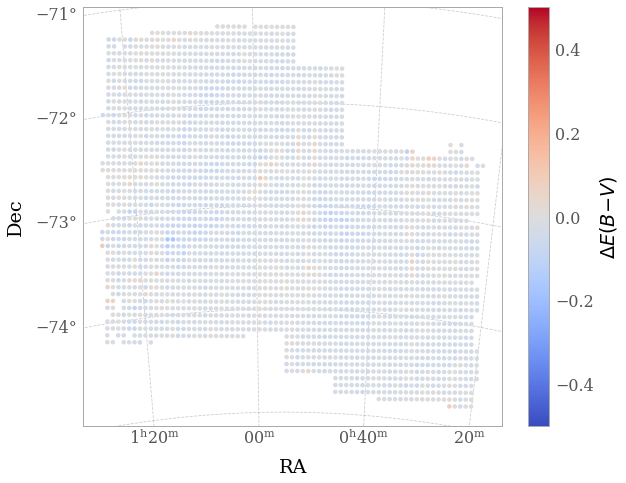}

    \includegraphics[width=0.9\linewidth]{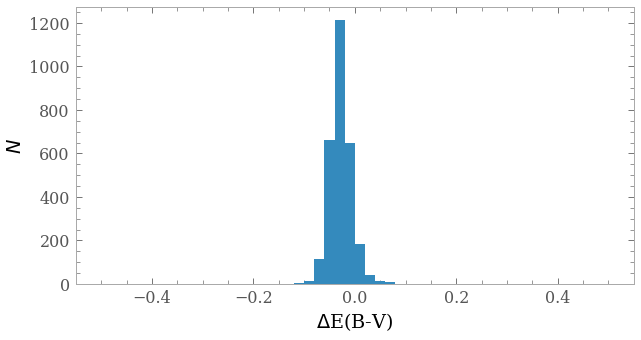}
    \caption{Comparison between our map for the SMC and the map by \cite{gorski2020}, but calibrated only with Na I D1-based reddening. Top: spatial distribution of differences. Bottom: histogram of differences.}
    \label{fig:comp_gorski_smc}
\end{figure}

\cite{gorski2020} used OGLE-III RC stars to construct LMC and SMC $E(B-V)$ maps. They reported mean reddening of 0.127 for the LMC and 0.08 for the SMC, which is higher than mean values in maps from this study, but note that their maps and maps from this study differ in coverage. \cite{gorski2020} calibrated the intrinsic RC color using several reddening tracers and methods, including reddening estimates derived from the Na I D1 absorption line, and adopted the mean value across all methods as the intrinsic color. For a direct comparison with our map, we used the version of their map calibrated exclusively with the Na I D1-based reddening. In Figures~\ref{fig:comp_gorski_lmc} and \ref{fig:comp_gorski_smc} we present the comparison between their maps and ours for the LMC and SMC, respectively. The differences appear as global shifts, with mean $\Delta E(B-V)$ values of $-0.05 \pm 0.03$ mag for the LMC and $-0.03 \pm 0.05$ mag for the SMC. Spatially, the strongest discrepancy manifests in the 30 Doradus region in the LMC, where the map by \cite{gorski2020} predicts higher reddening. Interestingly, when we compared our map with the version calibrated with the average of multiple reddening tracers, the 30 Doradus region did not stand out, as the map of differences contained many stronger, large-scale residuals that masked the 30 Doradus discrepancy. This demonstrates that different calibration approaches produce systematic differences between maps. Outside the 30 Doradus region, the version calibrated using Na I D1 shows the closest agreement with our map, suggesting that the Na I D1 absorption line provides a robust calibration for the RC intrinsic colour and serves as a reliable reddening tracer. The 30 Doradus region behaves anomalously in all RC-based comparisons, and we discuss this feature in detail later in the text.

\begin{figure}
    \centering
    \includegraphics[width=0.9\linewidth]{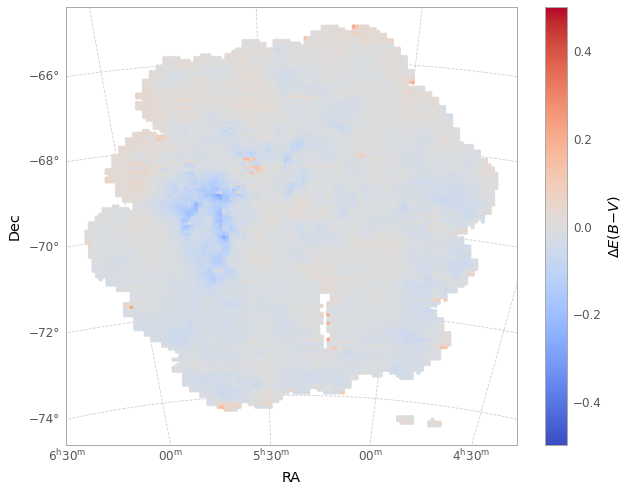}

    \includegraphics[width=0.9\linewidth]{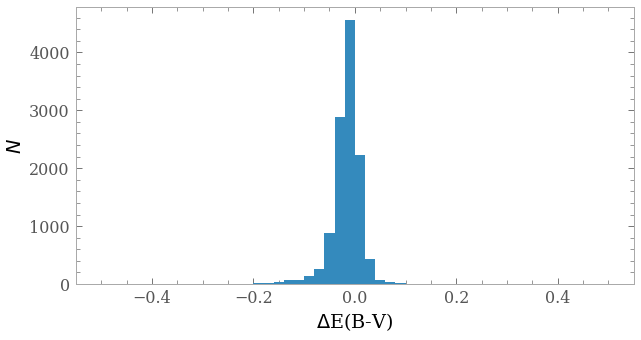}
    \caption{Comparison between our map for the LMC and the map by \cite{skowron2021}. Top: spatial distribution of differences. Bottom: histogram of differences.}
    \label{fig:comp_skowron_lmc}
\end{figure}

\begin{figure}
    \centering
    \includegraphics[width=0.9\linewidth]{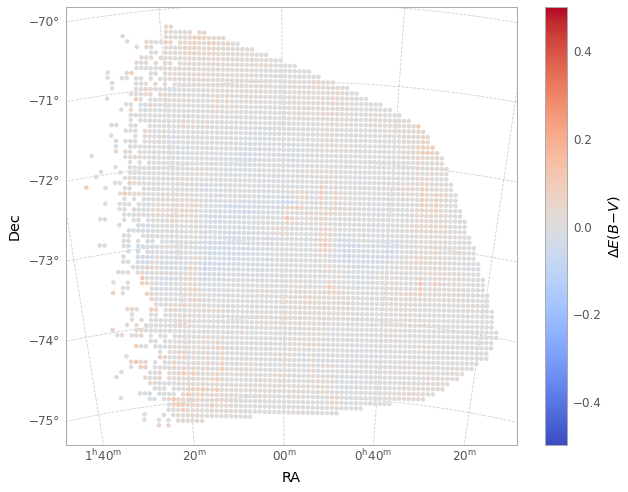}

    \includegraphics[width=0.9\linewidth]{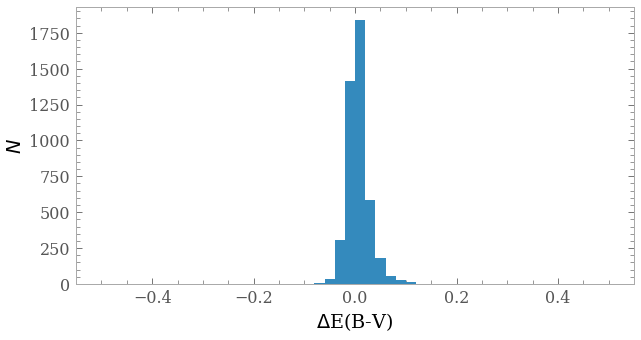}
    \caption{Comparison between our map for the SMC and the map by \cite{skowron2021}. Top: spatial distribution of differences. Bottom: histogram of differences.}
    \label{fig:comp_skowron_smc}
\end{figure}

\cite{skowron2021} calculated maps for LMC and SMC in $E(V-I)$. Using \texttt{dust\_extinction} \citep{gordon2024dust_extinction} and assuming the \cite{cardelliEtal1989} reddening law with $R_V=3.41$ we calculated the ratio $E(B-V)/E(V-I) = 1.313$, which was used to transorm their $E(V-I)$ values to $E(B-V)$ for the purpose of comparison. In Figures~\ref{fig:comp_skowron_lmc} and \ref{fig:comp_skowron_smc} we present comparison between our maps and results by \cite{skowron2021} for the overlapping area of the LMC and SMC, respectively. Mean $\Delta E(B-V)$ is $-0.016 \pm 0.029$ mag for the LMC, and $0.006 \pm 0.021$ mag for the SMC. The strongest discrepancy between the two maps is visible for the 30 Doradus region of the LMC, where our map predicts lower reddening than the map by \cite{skowron2021}.

\begin{figure}
    \centering
    \includegraphics[width=0.9\linewidth]{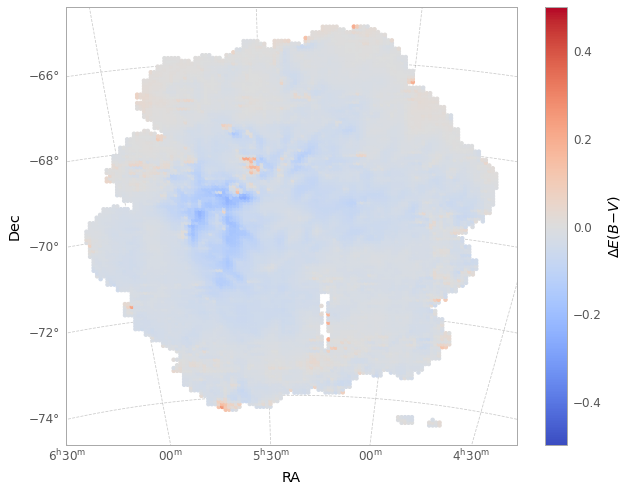}

    \includegraphics[width=0.9\linewidth]{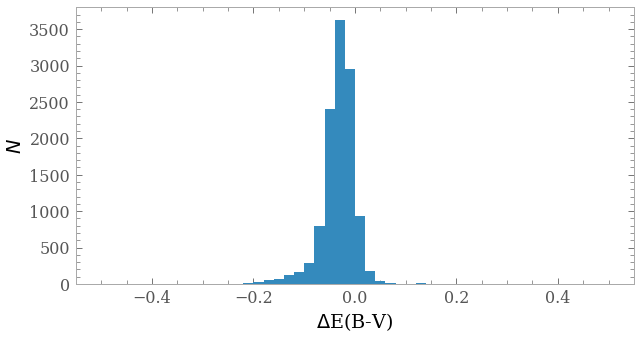}
    \caption{Comparison between our map for the LMC and the map by \cite{choi2018}. Top: spatial distribution of differences. Bottom: histogram of differences.}
    \label{fig:comp_choi_lmc}
\end{figure}

\cite{choi2018} used SMASH data to construct the reddening map of the LMC in $E(g-i)$ and reported the mean value of $E(g-i) = 0.15 \pm 0.05$ mag. Using $E(g-i)/E(B-V) = 1.724$, calculated under the same assumptions as above, we converted their map the $E(B-V)$ values and plot the differences in Fig.~\ref{fig:comp_choi_lmc}. Mean $\Delta E(B-V)$ is $-0.032 \pm 0.034$ mag. The largest difference between the two maps corresponds to the 30 Doradus region. Interestingly, this discrepancy is also visible when comparing our map to the maps based on RC color by \cite{gorski2020} and by \cite{skowron2021}, but no such discrepancy is present when comparing our map to other SED-based maps. This reveals that there is a difference between the two methods that manifest in this region. Our maps were computed assuming the \cite{gordon2003} reddening law with a single $R_V$ value applied across the entire galaxy. This assumption is inadequate specifically in the 30 Doradus region. The extinction law varies not only across the LMC and SMC but is different specifically in the 30 Doradus region \citep[e.g.][]{gordon2003}. Since the extinction law affects theoretical SEDs, this can lead to an underestimation of the reddening in our approach. On the other hand, RC intrinsic color might vary with environment due to its dependence on age and metallicity. Thus, in 30 Doradus, variations in the underlying stellar population may bias RC-based reddening estimates.

\section{Summary}\label{sec:summary}

We used the SED fitting method for multi-band $ugrizYJK_S$ photometry from the SMASH and VMC projects for RGB stars in order to construct reddening maps for the LMC and SMC. We investigated differences for three different sets of atmospheric models (K93, CK04, and Phoenix). 

Our resultant maps cover an area of 34.51 deg$^2$ for the LMC and 24.46 deg$^2$ for the SMC and both have 4 arcmin resolution. We compared our maps to selected literature studies which used SED fitting or RC star colors. Our findings are summarized below.

\begin{enumerate}
    \item The mean reddening values are $E(B-V)=0.076 \pm 0.022$ mag for the LMC and $E(B-V)=0.058 \pm 0.024$ mag for the SMC when using RGB stars. Maps based exclusively on RC stars yield nearly identical values and spatial distributions.
    \item Reddening distributions derived using Phoenix and CK04 models are consistent for both the LMC and SMC, while K93-based estimates are systematically shifted toward higher $E(B-V)$ values. This demonstrates that the choice of model atmosphere affects the absolute reddening scale, although relative spatial trends remain stable.
    \item Adopting the \cite{gordon2003} extinction law with the canonical values of $R_V=3.41$ for the LMC and $R_V=2.74$ for the SMC, we derived reddening estimates that are broadly consistent with previous photometric studies.   
    \item In addition to reddening, the fitting procedure explored a grid spanning effective temperatures, $\log g$, and [Fe/H]. While the latter two parameters were only coarsely sampled and were not analyzed further, the $T_{\rm eff}$ distributions provided a useful consistency check for the RGB population and are consistent with expectations for RGB, including slightly higher temperatures for the SMC sample due to its lower metallicity.
    \item Comparisons with other SED-based maps reveal only minor differences. The spatial distribution of differences show no distinct features or localized structures.
    \item Comparisons with RC-based maps show good agreement across most of the LMC and the SMC. A significant discrepancy appears only in 30 Doradus, likely driven by spatial variation in the reddening law and uncertainties in RC intrinsic colours in extreme environments; elsewhere, our maps agree well with RC-based maps.

\end{enumerate}

\begin{acknowledgements}
      This reasearch is funded by the European Research Council (ERC) under the European Union’s Horizon 2020 research and innovation program (grant agreement No. 951549 - UniverScale). We also acknowledge support from the Polish Ministry of Science and Higher Education grant 2024/WK/02. This research benefited from the Polish-French Marie Skłodowska-Curie and Pierre Curie Science Prize awarded by the Foundation for Polish Science. 
      This research has made use of the SVO Filter Profile Service "Carlos Rodrigo", funded by MCIN/AEI/10.13039/501100011033/ through grant PID2023-146210NB-I00. Herschel is an ESA space observatory with science instruments provided by European-led Principal Investigator consortia and with important participation from NASA. Based on data products created from observations collected at the European Organisation for Astronomical Research in the Southern Hemisphere under ESO programme 179.B-2003. This research made use of NASA’s Astrophysics Data System Bibliographic Services, as well as of the SIMBAD database operated at CDS, Strasbourg, France.
\end{acknowledgements}

\bibliographystyle{aa} 
\bibliography{bibliography} 

\begin{appendix}\label{sec:appendix-rc-stars}

\section{Reddening maps for K93 and CK04 models}\label{sec:other_models}

\begin{figure}[h]
\centering
\includegraphics[width=\linewidth]{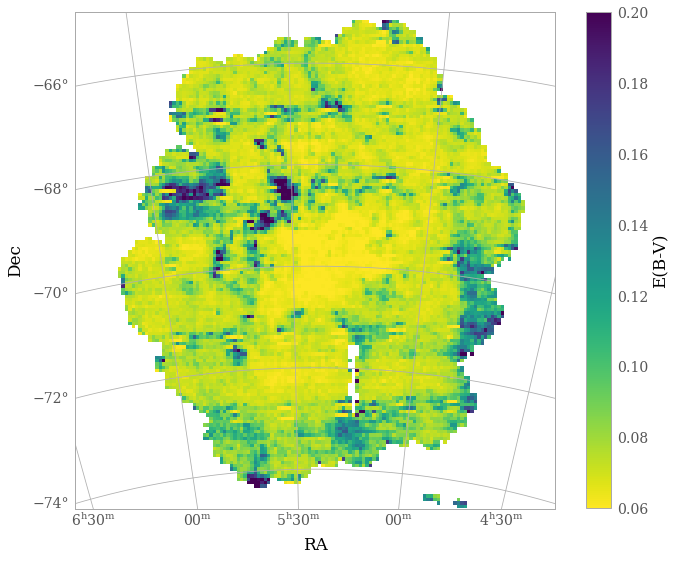}

\includegraphics[width=\linewidth]{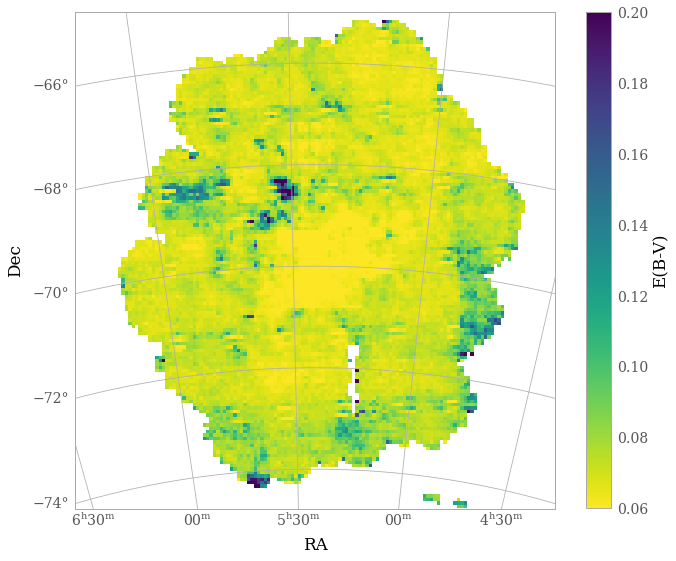}

\caption{Reddening maps for the LMC calculated with K93 (top) and CK04 (bottom) models.}
\label{fig:lmc_k93_ck04}
\end{figure}

\begin{figure}[h]
\centering
\includegraphics[width=\linewidth]{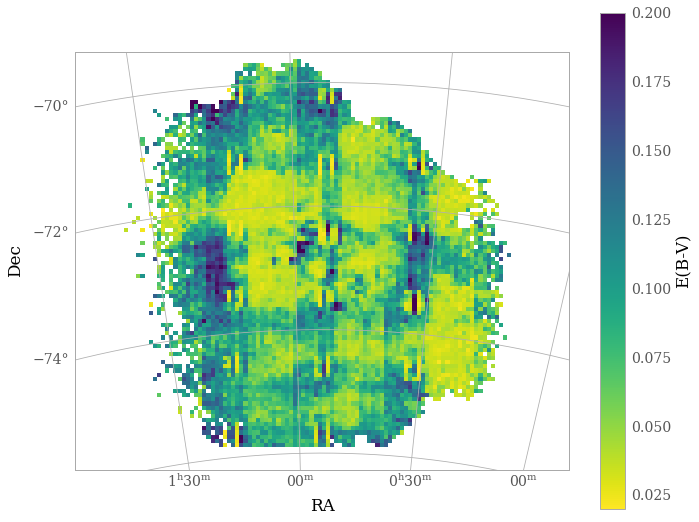}

\includegraphics[width=\linewidth]{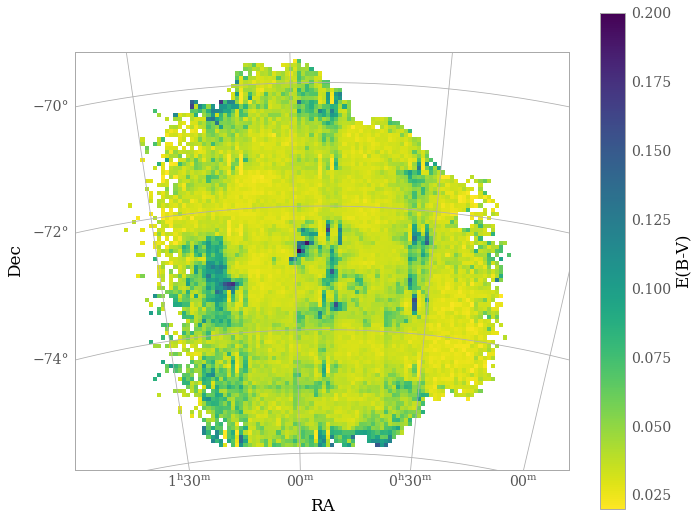}

\caption{Reddening maps for the SMC calculated with K93 (top) and CK04 (bottom) models.}
\label{fig:smc_k93_ck04}
\end{figure}

Here we present maps calculated using SEDs based on K93 and CK04 models. In Fig.~\ref{fig:lmc_k93_ck04} we plotted reddening maps for the LMC calculated with K93 and CK04 models. In Fig.~\ref{fig:smc_k93_ck04} we plotted reddening maps for the SMC calculated with K93 and CK04 models.

\end{appendix}
\end{document}